\documentstyle[twocolumn,prl,float,aps,epsfig,rotating]{revtex}
\begin{document}

\title{Ortho-para transition rate in $\mu$-molecular hydrogen
and the proton's induced pseudoscalar coupling $g_p$}

\author{
\mbox{J.H.D.~Clark$^{1}$}\cite{APSaddress}
\mbox{D.S.~Armstrong$^{1}$} 
\mbox{T.P.~Gorringe$^{2}$} 
\mbox{M.D.~Hasinoff$^{3}$}
\mbox{P.M.~King$^{1}$}\cite{ILLaddress}
\mbox{T.J.~Stocki$^{3}$}\cite{RPBaddress}
\mbox{S.~Tripathi$^{2}$}
\mbox{D.H.~Wright$^{4}$}\cite{SLACaddress}
\mbox{P.A.~Zolnierczuk$^{2}$}\cite{IUCFaddress}
}

\address{
\mbox{$^1$College of William and Mary, Williamsburg, VA USA 23185 }
\mbox{$^2$University of Kentucky, Lexington, KY 40506 }
\mbox{$^3$University of British Columbia, Vancouver, B.C., Canada V6T 1Z1 }
\mbox{$^4$TRIUMF, Vancouver, BC, Canada, V6T 2A3 }
}

\date{\today}
\maketitle

\begin{abstract}
We report a measurement of the ortho-para transition rate 
in the p$\mu$p molecule.
The experiment was conducted at TRIUMF
via the measurement 
of the time dependence
of the 5.2~MeV neutrons
from muon capture
in liquid hydrogen.
The measurement yielded an
ortho-para rate $\Lambda_{op} = 
(11.1 \pm 1.7 \pm^{0.9}_{0.6} ) \times 10^4$~s$^{-1}$
that is substantially larger 
than the earlier result
of Bardin {\it et al.}
We discuss the striking implications for the 
proton's induced pseudoscalar coupling $g_p$.
\end{abstract}


The protons's weak axial current
is governed by the weak axial form factor $F_A$($q^2$)
and the induced pseudoscalar form factor $F_P$($q^2$)
and their corresponding coupling constants $g_a = F_A (0)$
and $g_p = F_P ( -0.88 m_{\mu}^2)$.
The values of the coupling constants and the $q^2$ dependence of
the form factors are determined by the proton's sub-structure
and the underlying QCD dynamics.
The approximate chiral symmetry of QCD
implies a rigorous relationship 
between these constants that yields 
either $g_p = ( 6.5 \pm 0.2 )~g_a$ or 
$g_p = 8.2 \pm 0.2$ \cite{Go58,Be94,Fe97,Go04}, 
and its experimental verification 
is an important test of low energy QCD.

The dominant uncertainty in extracting $g_p$ from $\mu$ capture in
liquid H$_2$ is the ortho-para transition rate $\Lambda_{op}$ 
from the ortho ($\ell = 1$) $p \mu p$ state
to the para ($\ell = 0$) $p \mu p$ state.
This transition has been investigated theoretically
by several authors \cite{We60,Ba82,Fa99,Ge02}.
It proceeds by Auger emission
and its rate is a function of the electronic environment of 
the $p \mu p$ molecule, most notably analogues
of the neutral H$_1$ atom [$(p \mu p)^+ e$] 
and the charged H$^+_3$ ion [$(p \mu p)^+ H_2$]$^+$.
The only published measurement 
of the rate $(4.1 \pm 1.4)  \times 10^4$~s$^{-1}$
by Bardin {\it et al.}\ \cite{Ba81b}
is two standard deviations from the
calculated rate $(7.1 \pm 1.1)  \times 10^4$~s$^{-1}$ 
of Balakov {\it et al.}\ \cite{Ba82}.

Given its importance in determining $g_p$
we have performed a new measurement 
of the rate $\Lambda_{op}$.
Both the Bardin {\it et al.}\ experiment \cite{Ba81b} and the current experiment 
were based on the measurement of the time spectrum of the 5.2~MeV neutrons
from $\mu p \rightarrow n \nu$ capture.
Because of the spin dependence of the
capture rate, and the spin configurations
of the $\ell = 0, 1$ molecules,
the rate $\Lambda_{op}$ is encoded in the neutron time dependence. 
Neglecting the initial formation 
of ortho molecules,
the time dependence is
\begin{equation}
N(t) \propto e^{-\Lambda_D t} ( \Lambda_{PM}  + (\Lambda_{OM} - \Lambda_{PM} ) e^{- \Lambda_{op} t} ) ,
\label{e: n(t)}
\end{equation}
where 
$\Lambda_D$ is the muon disappearance rate
and $\Lambda_{OM}$ and $\Lambda_{PM}$ 
are the ortho and para capture rates.
Since  
$\Lambda_D$ 
and the hyperfine ratio 
$\Lambda_{OM} / \Lambda_{PM}$ 
are well established by experiment or theory,
the rate $\Lambda_{op}$ can be extracted
from the neutron time spectrum.

Our experiment was conducted on the M9B muon
channel at the TRIUMF cyclotron. 
Incoming muons were counted in a two-element
plastic scintillator beam telescope. 
The larger scintillator $S1$ (33$\times$33$\times$0.32~cm$^3$)
registered particles emerging from the beam pipe 
and the smaller scintillator $S2$ (10$\times$10$\times$0.16~cm$^3$) 
registered particles incident on the H$_2$ flask.
The target \cite{Wr92,Be95} comprised 
a cylindrical flask of length 15~cm and diameter 16~cm 
containing 2.7~liters of liquid H$_2$ 
with $4$$\pm$$2$~ppm deuterium contamination 
and  $<$$1$~ppb contamination from $Z$$>$$1$ impurities.
The target flask had Au front/side walls 
and a Cu back-plate
and the surrounding vacuum vessel had cylindrical Ag walls
and a mylar front window.

Outgoing neutrons were detected in a liquid scintillator counter array
consisting of four counters ($N2$-$N5$) with  diameters $12.7$~cm 
and thicknesses $12.7$~cm and one counter ($N1$) with a diameter $12.7$~cm and a thickness $5.1$~cm.
The scintillator materials were NE224 ($N1$), NE213 ($N4$) 
and BC501A ($N2$, $N3$, $N5$)
which permit n/$\gamma$ discrimination
by pulse-shape.
Sandwiched between the neutron counters and the vacuum vessel were  
plastic scintillator veto counters ($E1$-$E5$) for 
identification of $\mu$-decay electrons.

Signals from each neutron counter were fed to 
pulse-shape discrimination modules \cite{An87}. The
modules generated outputs indicating 
either a $\gamma$--ray trigger ($G$)
or a neutron trigger ($N$) 
via the comparison of the integrated signals in a 
420~ns prompt gate and a 52~ns delayed gate.
To reduce the neutron trigger rate from mis-identified 
$\gamma$--rays/electrons we rejected those triggers accompanied
by either a veto counter signal ($\overline{E}$)
or a large neutron counter signal ($\overline{O}$).
All events that fulfilled the neutron trigger
($N$$\cdot$$\overline{E}$$\cdot$$\overline{O}$)
and pre-scaled events that fulfilled the $\gamma$/e trigger 
($G$) were recorded. 
The pulse-shape data included 
both primary information from the pulse-shape modules (PSD1)
and secondary information from a charge integrating ADC (PSD2). 
Histories of beam telescope hits
and veto counter hits were recorded in a multihit TDC.

The experiment was conducted 
in four distinct data-sets and yielded a total of
3.7$\times$10$^{10}$ $\mu^-$ stops
in liquid H$_2$.
Empty target runs (yielding $0.9$$\times$$10^9$ $\mu^-$ stops)
and positive muon runs (yielding $2.0$$\times$$10^9$ $\mu^+$ stops) 
were used for background studies.
In sorting the data we applied cuts
to separate the  neutrons from $\mu$p capture,
with energies 5.2~MeV and lifetimes $\sim$2~$\mu$s,
from backgrounds which
included both mis-identified $\gamma$-rays
from electron bremsstrahlung 
and neutrons from nuclear $\mu$ capture, 
photo-nuclear reactions, accelerator sources, 
cosmic-rays, {\it etc}.

A pulse-shape cut and a pulse pile-up cut were used 
to minimize the contamination from mis-identified $\gamma$-rays.
The pulse-shape cut involved a comparison of the integrated
amplitudes in the signal peak and the signal tail, 
and used both PSD1 and PSD2 to achieve optimal 
n/$\gamma$ discrimination.
The pulse pile-up cut rejected any event with 
$\geq$1 neutron counter hits in a preceding 600~ns time window,
such pile-up potentially distorting the n/$\gamma$ discrimination.

An energy cut was used to enhance the signal of
recoil protons from 5.2~MeV neutrons
by requiring an energy equivalent of
typically $2.5$-$5.0$~MeV.
A blank cut was used to reject the events from
$\mu$ capture in high-Z materials 
({\it i.e.} $\tau_{Au} = 73$~ns,
$\tau_{Ag} = 87$~ns and $\tau_{Cu} = 163$~ns \cite{Su87})
by rejecting any event with $\geq$1~$S1$ hits
in a 400~ns window preceding
the neutron.
Finally, we used a duty cycle cut to reject any events 
occurring within 32~$\mu$s of the beam-off period in the cyclotron 
time structure, and an errant muon cut
to reject any events occurring
within 5~$\mu$s of an incoming beam particle 
that fired $S1$ but  missed $S2$.

\begin{figure}
\begin{center} 
\mbox{\epsfig{figure=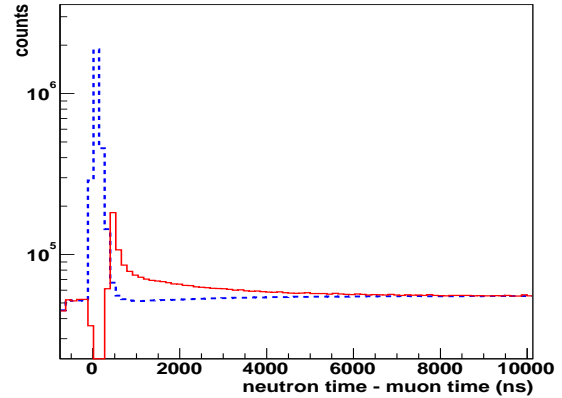,height=6cm,width=8cm}}
\end{center}
\caption{Time spectra between 
incoming muons and detected neutrons 
for events that passed all cuts {\it except} 
the blank cut (dashed line)
and all cuts {\it including} the blank cut (solid line).
The shallow hole for $t < 4$~$\mu$s that's visible
for events failing the blank cut
is due to random vetoing.
Note that $t =  -30$ to $370$~ns corresponds to the blank cut
and the dashed and solid curves are normalized
at $t > 10$~$\mu$s region.}
\label{fig1}
\end{figure}

The time spectra between incoming muons and detected neutrons 
for events that survived all cuts {\it excluding}
the blank cut and all cuts {\it including}
the blank cut are plotted in Fig.\ \ref{fig1}.
Note that for each event we increment the time spectrum
for every $S1$$\cdot$$S2$ coincidence
in the multihit TDC.
The spectra show
a short-lifetime ($\tau \sim 80$~ns) component 
that is most evident for events that failed
the blank cut, 
a long-lifetime ($\tau \sim 2$~$\mu$s) component 
that is only evident for events that
passed the blank cut,
and a continuum background. 
The short-lifetime component is dominated by Ag/Au capture,
the  long-lifetime component is dominated by H$_2$ capture,
and the continuum arises from uncorrelated muons. 

The procedures for accounting for backgrounds 
in the time spectrum of the events passing all cuts 
were (i) subtraction 
of the continuum background from uncorrelated
muons,  (ii) determination of contributions
from photo-nuclear reactions following
electron bremsstrahlung,
and (iii) the determination of contributions 
from $\mu$ capture in low-Z materials.

The continuum
is not perfectly time independent as decay electrons from
uncorrelated muons can reject events by firing 
the veto counter ($\overline{E}$)
or overload circuit ($\overline{O}$).
This random vetoing of neutron events by uncorrelated muons
can be seen as the shallow hole for $t < 4$~$\mu$s 
for events failing 
the blank cut.
Note that for times $t >> \tau_{Au/Ag}$ 
the events failing the blank cut are entirely
dominated by the continuum background
while the events passing the blank cut
are comprised of $\mu$p signal
and continuum background.
Thus, by normalizing and subtracting the events
that failed the blank cut from those that passed the 
blank cut, the continuum
is removed.

The photo-nuclear background originates from the
photo-nuclear interactions [{\it e.g.} $^{12}$C($\gamma $,$p$)] 
in the liquid scintillators. 
Consequently, a tiny fraction
of bremsstrahlung photons 
from decay electrons
yield proton recoils.
This background carries the 
2.2~$\mu$s lifetime of muonic H$_2$
and therefore distorts the 
effective disappearance rate 
of $\mu$p capture neutrons.
To measure this background
we conducted a $\mu^+$ measurement
where neutrons from $\mu$ capture are absent 
but those from photo-nuclear reactions are present.
From the amplitude of the $2.2$~$\mu$s lifetime component 
in the $\mu^+$ neutron time spectrum
we determined a contribution of photo-nuclear interactions
to $\mu^-$ running that averaged about $14$\% 
for the thick counters N$2$-N$5$
but rose to $55$$\pm$8\% for 
the thin counter $N1$.
We used a GEANT3 simulation \cite{GEANT} to account for 
the difference in the $\gamma$-ray spectra from 
$\mu^+$ and $\mu^-$ stops due to $e^+$ annihilation.

The background from $\mu$ capture in low-Z materials,
{\it e.g.} scintillators, light-pipes,
is worrisome as the muonic C and H$_2$ lifetimes 
are similar (2.028~$\mu$s and 2.195~$\mu$s).
This background was determined by performing 
an empty target run
in which $\mu^-$ stops
occur in the Cu back-plate
not the liquid H$_2$.
A GEANT3 simulation
was used
to establish that stops in scintillators, light guides, {\it etc.}, 
were dominated by scattering from the beam pipe window
and the beam telescope
and thus unaffected by the target filling.
The measurement showed
no evidence for backgrounds from low-Z capture 
and established a limit of
$<$7\% on carbon backgrounds
in production running.

\begin{figure}
\begin{center} 
\mbox{\epsfig{figure=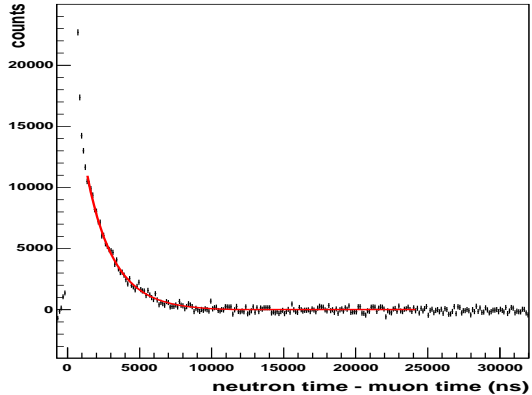,height=6cm,width=8cm}}
\end{center}
\caption{Time spectrum of events that survive all cuts
after the continuum subtraction.
The points are the data and the solid curve is 
the benchmark fit.}
\label{fig3}
\end{figure}

\begin{figure}
\begin{center} 
\mbox{\epsfig{figure=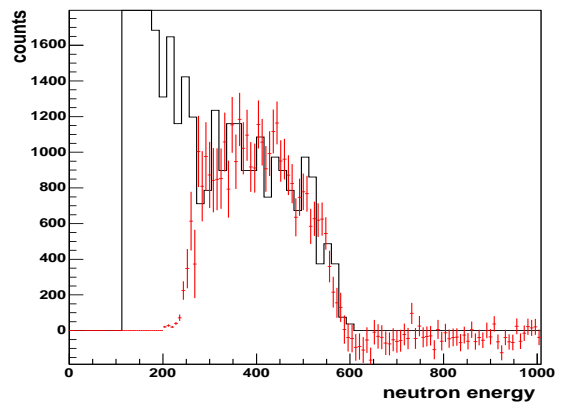,height=6cm,width=8cm}}
\end{center}
\caption{Energy spectrum for counter $N4$ and data-set I
for events surviving all cuts with
times $t = 1.1$-$3.3$~$\mu$s
after continuum subtraction. 
The points
are the data and the histogram is 
the simulation  The data's upper edge (chan.\ $\sim$550)
corresponds to 5.2~MeV
proton recoils and the data's lower edge (chan.\ $\sim$250)
results from the pulse-shape cut.}
\label{fig4}
\end{figure}

The time spectrum of neutron events that survived all cuts,
after the subtraction of the continuum, is
plotted in Fig.\ \ref{fig3}.
A typical energy spectrum ($N4$, data-set I)
is plotted in Fig.\ \ref{fig4}.
It shows the expected spectrum
for 5.2~MeV neutrons as obtained by
a GEANT3 simulation of the experimental set-up.

When muons stop in liquid H$_2$
they evolve through a sequence of different atomic and molecular
states. 
Also, if D$_2$ is present
other species of $\mu$-atoms and $\mu$-molecules are 
generated \cite{Wa65,Pe90}.
Therefore corrections were applied to Eqn.\ \ref{e: n(t)}
to account for the p$\mu$p formation from 
the singlet $\mu$p atoms
and the p$\mu$d formation
due to D$_2$ contamination.
We assumed that, due to their typically lower energies,
the efficiency for detecting neutrons
from d/$^3$He capture is significantly smaller
than the efficiency for detecting neutrons
from proton capture,
this making proton capture in p$\mu$d molecules
the dominant correction from D$_2$ contamination.

The determination of $\Lambda_{op}$
was obtained by fitting
to the Fig.\ \ref{fig3} time spectrum.
The fitting function was based on Eqn.\ \ref{e: n(t)}, 
but supplemented with corrections accounting 
for p$\mu$p formation and
D$_2$ contamination.
We also included a correction term with 
disappearance rate $\Lambda_D$
to account for photo-nuclear backgrounds.
In most fits we fixed the ratio 
$\Lambda_{OM}$/$\Lambda_{PM} = 2.59$ \cite{Be94,Fe97} and rate 
$\Lambda_D = 4.55 \times 10^5$~s$^{-1}$ \cite{Ba81a} at their known values 
and only varied the overall normalization 
and rate $\Lambda_{op}$.

Our benchmark fit (fit A) to all counters and all
data-sets, which demonstrates the good agreement
between the measured spectrum and the theoretical spectrum, 
is shown in Fig.\ \ref{fig3}.
This yielded a rate 
$\Lambda_{op}$ = $ ( 11.1 \pm 1.7 ) \times 10^4$~s$^{-1}$ 
with a chi-squared $p.d.f.$ of $1.03$
for a fit range of $t = 1.3$ 
to $24.3$~$\mu$s, {\it i.e.}
beginning beyond the region contaminated by
Au/Ag capture.

The results of fits to study
the sensitivity to
backgrounds and corrections
are listed in Table \ref{table1}.
It shows that omitting the p$\mu$p formation correction (fit B)
or p$\mu$d contamination (fit C)
correction has little effect ($< 1$~$\sigma$) 
on $\Lambda_{op}$.
Including a 7\% carbon background (fit D) changed the
rate from $( 11.1 \pm 1.7 ) \times 10^4$~s$^{-1}$
to $( 11.7 \pm 2.0 ) \times 10^4$~s$^{-1}$,
a $0.5$~$\sigma$ effect,
while omitting the photo-nuclear background (fit E)
changed the rate from $( 11.1 \pm 1.7 ) \times 10^4$~s$^{-1}$
to $( 8.0 \pm 1.1 ) \times 10^4$~s$^{-1}$,
a $1.8$~$\sigma$ effect.
Other tests indicated that varying 
the hyperfine ratio $\Lambda_{OM}$/$\Lambda_{PM}$
by $\pm$5\% changed the ortho-para rate $\Lambda_{op}$ by $\pm$4\%.

Also given in Table \ref{table1} are results from fits 
to the individual time spectra of the five neutron counters,
thus demonstrating the consistency between counters.
Similar checks of separate fits to the
four different data-sets from the two running periods, which involved 
some differences in calibrations and settings,
yielded results ranging from 
$\Lambda_{op} = ( 10.3 \pm 2.6 ) \times 10^4$~s$^{-1}$ to 
$( 13.7 \pm 3.0 ) \times 10^4$~s$^{-1}$.
Lastly, we found no evidence of significant sensitivity 
to either the start-time or the end-time of the fit.

Several other tests of data integrity were performed.
We sub-divided the neutron events
into a low energy data-set (typically $2.5$-$3.7$~MeV) and 
a high energy data-set (typically $3.7$-$5.0$~MeV) 
and found consistent results for $\Lambda_{op}$. 
We analyzed the neutron events with energies $> 5.2$~MeV 
and found no evidence 
for a long-lifetime component.
Finally, the time spectrum of electron
events yielded $\tau = 2.198 \pm 0.002$~$\mu$s and
the time spectrum of empty-target neutron events
yielded $\tau = 163 \pm 1$~ns, consistent with the known values
for the $\mu$p lifetime  \cite{Ba81a} and the $\mu$Cu lifetime
\cite{Su87}.

Our final result is $\Lambda_{op} = ( 11.1 \pm 1.7 \pm^{0.9}_{0.6} ) 
\times 10^4$~s$^{-1}$  where the first error is statistical 
and the second error accounts for uncertainties
in background determinations,
chemistry corrections 
and $\Lambda_{OM}$/$\Lambda_{PM}$ ($\pm$5\%).
Our result is somewhat larger (by 1.9~$\sigma$)
than the theoretical estimate $\Lambda_{op} 
= ( 7.1 \pm 1.1 ) \times 10^5$~s$^{-1}$
of  Bakalov {\it et al}.\  \cite{Ba82}
and substantially larger (by 3.1~$\sigma$) than the
earlier measurement
$\Lambda_{op} = ( 4.1 \pm 1.4 ) \times 10^4$~s$^{-1}$
of Bardin {\it et al}.\  \cite{Ba81b}.

In extracting $g_p$ from $\mu$ capture in liquid H$_2$
the difference between $\Lambda_{op} = 4.1 \times 10^4$~s$^{-1}$
and $11.1 \times 10^4$~s$^{-1}$ is striking.
With the Bardin {\it et al.}\ value
$\Lambda_{op} = 4.1 \times 10^4$~s$^{-1}$
the TRIUMF RMC experiment \cite{Jo96,Wr98} yields $g_p = 12.2 \pm 1.1$ \cite{Jo96,Wr98},
the Saclay OMC experiment \cite{Ba81a} yields $g_p = 10.6 \pm 2.7$ \cite{Ba81a},
and the average value of the earlier liquid H$_2$ experiments is $g_p =  10.4 \pm 4.1$ \cite{Bl62,Ro63},
with the TRIUMF result being clearly inconsistent
with theory.
With the new measured value $\Lambda_{op} = 11.1 \times 10^4$~s$^{-1}$
the TRIUMF RMC experiment yields $g_p = 10.6 \pm 1.1$,
the Saclay OMC experiment yields $g_p = 0.8 \pm 2.7$,
and the average value of the earlier liquid H$_2$ experiments yields $g_p = 5.6 \pm 4.1$,
with the Saclay result being clearly inconsistent
with theory.

The inconsistency between the present value for $\Lambda_{op}$
and that measured by Bardin {\it et al.}\ 
is disconcerting.
Note that an important difference between the two experiments 
is the beam time structure, the
earlier work using a
pulsed beam while the current work using a
continuous beam.
The experiments also differed in the total number of 5.2~MeV neutrons,
the signal-to-noise for 5.2~MeV neutrons,
and the presence in the Bardin {\it et al.}\ experiment 
of a long-lifetime background.
However, we see no obvious explanation for the 
conflicting results.

Note that one consequence of the different 
time structures in the two experiments
is the typical time between the parent muon and the detected neutron,
in the relevant region of the time spectra,
is greater in the Bardin {\it et al.}\ experiment
than the current experiment.
Therefore, at least in principle, it is possible 
that unconventional $\mu$-chemistry in liquid hydrogen  
could manifest itself as
different neutron disappearance rates in the
two experiments.

In summary, we report a determination of the $p \mu p$ ortho-para rate
from the time spectrum of the 5.2 MeV $\mu p$ 
capture neutrons.
Our result 
$\Lambda_{op} = ( 11.1 \pm 1.7 \pm^{0.9}_{0.6} ) \times 10^4$~s$^{-1}$
is somewhat larger than the theoretical result
of Bakalov {\it et al.}\ \cite{Ba82} and strikingly larger
than the only previously published measurement \cite{Ba81b}.
The new result has dramatic consequences for the determination 
of the induced pseudoscalar coupling $g_p$.

We thank the TRIUMF cyclotron staff 
and TRIUMF target group  for the operation 
of the cyclotron and the target.
We also thank 
the National Science Foundation (USA),
Natural Sciences and Engineering Research Council (Canada),
and Jeffress Memorial Trust and
the William \& Mary Endowment
for their financial support.

\vspace{-0.3cm}

\begin{table}
\caption{Results for the rate $\Lambda_{op}$
from  the fits to the continuum-subtracted 
time spectra. Values are given for both the summed spectrum 
of all neutron counters and the individual spectra of
each neutron counter.
See text for explanation of 
fits A,B, C, D and E.
The units of $\Lambda_{op}$ are $\times 10^4$~s$^{-1}$.}
\label{table1}
\begin{tabular}{ccccccc}
 Cntr.\ & Fit A. & Fit B & Fit C & Fit D & Fit E \\ 
\hline
All & 11.1$\pm$1.7 & 12.2$\pm$1.8 & 11.6$\pm$1.7 &  11.7$\pm$2.0 &  8.0$\pm$1.1 \\
N1  & 7.2$\pm$10.6 & 8.2$\pm$11.1 & 7.7$\pm$10.6 &  7.3$\pm$10.9 &  2.2$\pm$3.9 \\
N2  & 13.8$\pm$5.5 & 15.4$\pm$5.6 & 14.4$\pm$5.5 &  14.8$\pm$6.5 &  8.3$\pm$3.1 \\
N3  &  8.9$\pm$2.1 &  9.9$\pm$2.3 &  9.4$\pm$2.2 &  9.2$\pm$2.3 &  7.1$\pm$1.7 \\
N4  & 14.6$\pm$6.2 & 16.1$\pm$6.7 & 15.1$\pm$6.2 &  15.5$\pm$7.3 & 10.3$\pm$4.1 \\
N5  & 12.2$\pm$3.0 & 13.5$\pm$3.0 & 12.7$\pm$3.0 &  12.9$\pm$3.4 & 10.6$\pm$2.6 \\
\end{tabular}
\end{table}


\vspace{-0.4cm}

\begin{thebibliography}{99}

\vspace{-01.5cm}

\bibitem[\dag]{SLACaddress} Present address: 
SLAC, P.O. Box 20450, Stanford, CA 94309.

\bibitem[\ddag]{APSaddress}Present address:
American Physical Society, One Physics Ellipse, College Park, MD 20740.

\bibitem[\star]{IUCFaddress}Present address:
Indiana University Cyclotron Facility,
2401 Milo B Sampson Lane,
Bloomington,IN 47408.

\bibitem[\circ]{ILLaddress} Present address:
University of Illinois at Urbana-Champaign,
Urbana, IL, 61801.

\bibitem[\diamond]{RPBaddress} Present address:
Radiation Protection Bureau, Health Canada, 775 Brookfield Road, 
Ottawa, ON, Canada.

\vspace{0.2cm}

% gp theory

\bibitem{Go58}
M.L.~Goldberger and S.B.~Treiman, 
Phys.\ Rev.\ {\bf 111}, 354 (1958).

\bibitem{Be94}
V.~Bernard, N.~Kaiser and Ulf-G.~Meissner, 
Phys.\ Rev.\ D {\bf 50}, 6899 (1994).

\bibitem{Fe97}
Harold~W.~Fearing, Randy Lewis, Nader Mobed, and Stefan Scherer, 
Phys.\ Rev.\ D {\bf 56}, 1783 (1997).

\bibitem{Go04}
T.P.~Gorringe and H.W.~Fearing,
Rev.\ Mod.\ Phys.\  {\bf 76}, 31 (2004).

% ortho-para theory

\bibitem{We60}
Steven Weinberg, 
Phys.\ Rev.\ Lett.\ {\bf 4}, 575 (1960).

\bibitem{Ba82}
D.D.~Bakalov {\it et al.}, 
Nucl.\ Phys.\ A {\bf 384}, 302 (1982).

\bibitem{Fa99}
M.P.~Faifman and L.I.~Men'shikov, 
Hyperfine Interactions {\bf 118}, 187 (1999).

\bibitem{Ge02}
S.S.~Gershtein and A.V.~Luchinsky,
Phys.\ of Atomic Nuclei {\bf 65}, 102 (2002).

% ortho-para expt

\bibitem{Ba81b}
G.~Bardin {\it et al.},
Phys.\ Lett.\ B {\bf 104}, 320 (1981).

% LH2 target

\bibitem{Wr92} D.~Wright {\em et~al.}, 
Nucl.\ Instrum.\ Methods {\bf A320}, 249 (1992).

\bibitem{Be95} W.~Bertl {\it et al.},
Nucl.\ Instrum.\ Methods Phys.\ Res.\ 
{\bf A 355}, 230 (1995).

% pulse-shape electronics

\bibitem{An87}
J.R.M.~Annand, Nucl.\ Instrum.\ and Methods {\bf A 262}, 371, (1987).

% muon disappearance rates

\bibitem{Su87} T. Suzuki, D.F. Measday, and J.P. Roalsvig,  Phys.
Rev. {\bf C 35}, 2212 (1987).

% GEANT

\bibitem{GEANT} R.~Brun {\it et al.}\
GEANT3(1986); CERN report no. DD/EE/84--1 (unpublished).

% deuterium effects

\bibitem{Wa65}
I.T.~Wang {\it et al}.\ 
Phys.\ Rev.\ {\bf 139}, B1528 (1965).

\bibitem{Pe90}
C.~Petitjean {\it et al.},
Muon Catal.\ Fusion {\bf 5/6}, 199 (1990/91).

% OMC/RMC on H expt

\bibitem{Jo96}
G.~Jonkmans {\it et al.}, 
Phys.\ Rev.\ Lett.\ {\bf 77}, 4512 (1996).

\bibitem{Wr98}
D.~H.~Wright, {\it et al.}, 
Phys.\ Rev.\ C  {\bf 57}, 373 (1998).


\bibitem{Ba81a}
G.~Bardin {\it et al.},
Nucl.\ Phys.\ A {\bf 352}, 365, (1981).


\bibitem{Bl62}
E.~Bleser {\it et al.},
Phys.\ Rev.\ Lett. {\bf 8}, 288 (1962).

\bibitem{Ro63}
J.~E.~Rothberg {\it et al.}\
Phys.\ Rev.\ {\bf 132}, 2664 (1963).


\end{thebibliography}
\end{document}